\documentclass[prb,twocolumn,showpacs,preprintnumbers,amsmath,amssymb,superscriptaddress, bibnotes]{revtex4}

\usepackage{color}
\usepackage{graphicx}
\usepackage{dcolumn}
\usepackage{bm}

\begin{document}
\newcommand{\red}[1]{\textcolor{red}{#1}}

\title{Magnetic and superconducting properties on {\it S}-type single-crystal CeCu$_2$Si$_2$ probed by $^{63}$Cu nuclear magnetic resonance and nuclear quadrupole resonance}

\author{Shunsaku~Kitagawa}
\email{kitagawa.shunsaku.8u@kyoto-u.ac.jp}
\author{Takumi~Higuchi}
\author{Masahiro~Manago}
\author{Takayoshi~Yamanaka}
\author{Kenji~Ishida}
\email{kishida@scphys.kyoto-u.ac.jp}
\affiliation{Department of Physics, Kyoto University, Kyoto 606-8502, Japan}
\author{H.~S.~Jeevan}
\author{C.~Geibel}
\affiliation{Max-Planck Institute for Chemical Physics of Solids, D-01187 Dresden, Germany}

\date{\today}

\begin{abstract}
We have performed $^{63}$Cu nuclear magnetic resonance/nuclear quadrupole resonance measurements to investigate the magnetic and superconducting (SC) properties on a ``superconductivity dominant'' ($S$-type) single crystal of CeCu$_2$Si$_2$. 
Although the development of antiferromagnetic (AFM) fluctuations down to 1~K indicated that the AFM criticality was close, Korringa behavior was observed below 0.8~K, and no magnetic anomaly was observed above $T_{\rm c} \sim$ 0.6 K.
These behaviors were expected in $S$-type CeCu$_2$Si$_2$.  
The temperature dependence of the nuclear spin-lattice relaxation rate $1/T_1$ at zero field was almost identical to that in the previous polycrystalline samples down to 130~mK, but the temperature dependence deviated downward below 120~mK.
In fact, $1/T_1$ in the SC state could be fitted with the two-gap $s_{\pm}$-wave rather than the two-gap $s_{++}$-wave model down to 90~mK.
Under magnetic fields, the spin susceptibility in both directions clearly decreased below $T_{\rm c}$, indicative of the formation of spin singlet pairing.
The residual part of the spin susceptibility was understood by the field-induced residual density of states evaluated from $1/T_1T$, which was ascribed to the effect of the vortex cores.
No magnetic anomaly was observed above the upper critical field $H_{c2}$, but the development of AFM fluctuations was observed, indicating that superconductivity was realized in strong AFM fluctuations.  
\end{abstract}

\maketitle
\section{Introduction} 
Since the discoveries of unconventional superconductivity in heavy-fermion (HF)\cite{F.Steglich_PRL_1979,H.R.Ott_PRL_1983,G.R.Stewart_PRL_1984,C.Pfleiderer_RMP_2009}, organic\cite{D.Jerome_JPhysLett_1980,K.Bechgaard_PRL_1981}, and cuprate compounds\cite{J.G.Bednorz_ZPB_1986,C.W.Chu_PhysicaB_1987,M.K.Wu_PRL_1987}, many studies have attempted to elucidate the pairing mechanism of these superconductors.
Identifying the superconducting (SC) gap structure is one of the most important issues since the SC gap structure is closely related to the SC pairing mechanism.
In particular, $k$-dependent pairing interactions lead to non-$s$-wave symmetry in unconventional superconductors. Among the HF superconductors, the pairing symmetry of CeCoIn$_{5}$ has been identified to be $d_{x^2-y^2}$-wave from field-angle-resolved experiments\cite{K.Izawa_PRL_2001,K.An_PRL_2010} and scanning tunneling microscopy measurements\cite{M.P.Allan_NatPhys_2013}; thus the superconductivity is considered to be mediated by antiferromagnetic (AFM) fluctuations as in the case of the cuprate superconductivity.

The first HF superconductor discovered in 1979\cite{F.Steglich_PRL_1979}, CeCu$_2$Si$_2$, was also considered to be a nodal unconventional superconductor since the SC phase was located on the verge of the AFM phase. Moreover, the $T^3$ dependence of the nuclear spin-lattice relaxation rate $1/T_1$, together with the absence of a coherence peak\cite{Y.Kitaoka_JPSJ_1986,K.Ishida_PRL_1999,K.Fujiwara_JPSJ_2008} and the $T^2$-like temperature dependence of the specific heat\cite{J.Arndt_PRL_2011} in the SC state, indicated a line nodal SC gap in CeCu$_2$Si$_2$. 
Finally, a clear spin excitation gap was observed in the SC state with inelastic neutron scattering, suggesting that AFM fluctuations were the main origin of superconductivity in CeCu$_2$Si$_2$\cite{O.Stockert_PhysicaB_2008,O.Stockert_NaturePhys_2011}. 
The clear decrease of the nuclear magnetic resonance (NMR) Knight shift below $T_{\rm c}$\cite{Y.Kitaoka_JJAP_1987} and the strong limit of the upper critical field $H_{\rm  c2}$\cite{H.A.Vieyra_PRL_2011}, plausibly originating from the Pauli-paramagnetic effect, indicated that the SC pairs were singlets. 
These results were considered to be evidence of a $d$-wave gap symmetry with line nodes in CeCu$_{2}$Si$_{2}$, such as a $d_{x^2-y^2}$- or $d_{xy}$-wave.

One difficulty in studying CeCu$_2$Si$_2$ is that a stoichiometric CeCu$_2$Si$_2$ is located very close to a magnetic quantum critical point, resulting in a ground state that is quite sensitive to the actual stoichiometry\cite{F.Steglich_PhysicaB_1996,S.Serio_PSPB_2010}. 
After careful sample-dependence experiments as well as experiments with chemical (Ge-substitution) and hydrostatic pressures, the ground state of the stoichiometric CeCu$_2$Si$_2$ was found to be the SC state coexisting with an unusual magnetic state called an ``$A$'' phase\cite{K.Ishida_PRL_1999,R.Feyerherm_PRB_1997,E.Vargoz_JMMM_1998,H.Q.Yuan_Science_2003}.
In this coexisting ``$A/S$'' sample, superconductivity expels the magnetic $A$ phase below $T_{\rm c}$ and becomes dominant at $T \rightarrow 0$\cite{R.Feyerherm_PRB_1997}.
The ground state of the $A$ phase was unclear for a long time. The ground state was revealed by elastic neutron scattering with the $A$-type single-crystal CeCu$_2$Si$_2$\cite{O.Stockert_PRL_2004}, and 
the nature of the $A$ phase was shown to be a spin-density-wave (SDW) instability from the observation of long-range incommensurate AFM order. 
Thus, an SC sample that does not show the $A$-phase behavior is located at the Cu-rich side, e.g., CeCu$_{2.2}$Si$_2$, which is called an ``$S$''-type sample.
 
Another difficulty in studying CeCu$_2$Si$_2$ is that large single-crystal samples showing superconductivity were not available before 2000, and thus, most measurements were performed on well-characterized polycrystalline samples.
Consequently, axial-dependent and angle-resolved measurements have not been performed.
However, large single crystals with well-defined properties have been synthesized and have recently been used for various experiments. 
In particular, recent specific-heat measurements on an $S$-type CeCu$_2$Si$_2$ single crystal down to 40~mK strongly suggested that CeCu$_2$Si$_2$ possesses a full gap with a multiband character\cite{S.Kittaka_PRL_2014}.
In addition, the small $H$-linear coefficient of the specific heat at low temperatures and its isotropic $H$-angle dependence under a rotating magnetic field within the $ab$ plane sharply contrast the expected behaviors in nodal $d$-wave superconductivity. 

In this study, we have performed $^{63}$Cu-NMR/nuclear quadrupole resonance (NQR) measurements to investigate the SC and magnetic properties of an $S$-type single crystal of CeCu$_2$Si$_2$.
As far as we know, this is the first NMR/NQR measurements on a single-crystal CeCu$_2$Si$_2$ down to 90~mK.
Comparison between the NMR results of previous polycrystalline and single-crystal samples is very important to understand the nature of superconductivity in CeCu$_2$Si$_2$.  
We found that the temperature dependence of $1/T_1$ at zero field was almost the same as that in previous polycrystalline $S$- and $A/S$-type samples down to 130~mK, but deviated downward below 120~mK.
The $T$ dependence of $1/T_1$ down to 90~mK could be reproduced by the two-gap $s_{\pm}$-wave model and the two-band $d$-wave model.
Taking into account the recent results of the field-angle dependence of the specific heat, the two-gap $s_{\pm}$-wave model is plausible.
The Knight shift parallel and perpendicular to the $c$-axis decreased in the SC state, in good agreement with previous results.
The magnitude of the residual Knight shift was analyzed with the $1/T_1$ result in magnetic fields and was ascribed to the field-induced density of states originating from the vortex effect.
In addition, we also investigated whether magnetic ordering was observed above the upper critical magnetic field $H_{\rm c2}$ since this anomaly was reported above $H_{\rm c2}$ with magnetoresistance and de Haas--van Alphen measurements\cite{H.Nakamura_JMMM_1988,F.Steglich_JPCM_1989,M.Hunt_JPCM_1990}.
No magnetic ordering was observed in the present $S$-type single crystal, but the development of AFM fluctuations was observed.

\section{Experimental}
Single crystals of CeCu$_{2}$Si$_{2}$ were grown by the flux method\cite{S.Serio_PSPB_2010}.
In the present NMR/NQR measurements, we used high-quality $S$-type single crystals from the same batch as those used in the specific-heat and magnetization measurements\cite{S.Kittaka_PRL_2014,S.Kittaka_PRB_2016}.
A single-crystal sample was used for NQR measurements without being powdered, and the NQR results of the single crystal were compared with the previous results measured in polycrystalline samples. 
Low-temperature NMR/NQR measurements were carried out with a $^3$He - $^4$He dilution refrigerator, in which the sample was immersed into the $^3$He - $^4$He mixture to avoid rf heating during measurements. The external fields were controlled by a single-axis rotator with an accuracy better than 0.5$^{\rm o}$.
The $^{63}$Cu-NMR/NQR spectra (nuclear spin $I$ = 3/2, and nuclear gyromagnetic ratio $^{63}\gamma/2\pi = 11.285$~MHz/T) were obtained as a function of frequency in a fixed magnetic field.
The NMR measurements were done at $\mu_0H \sim 1.4$~T ($< \mu_0H_{\rm c2} \sim 2$~T) and $\sim 3.5$~T ($> \mu_0H_{\rm c2}$).
The $^{63}$Cu Knight shift of the sample was calibrated by the $^{63}$Cu signals from the NMR coil.
The $^{63}$Cu nuclear spin-lattice relaxation rate $1/T_1$ was determined by fitting the time variation of the spin-echo intensity after saturation of the nuclear magnetization to a theoretical function for $I$ = 3/2\cite{A.Narath_PR_1967,D.E.MacLaughlin_PRB_1971}.

\section{Experimental Results}
\begin{figure}[!tb]
\vspace*{10pt}
\begin{center}
\includegraphics[width=8.5cm,clip]{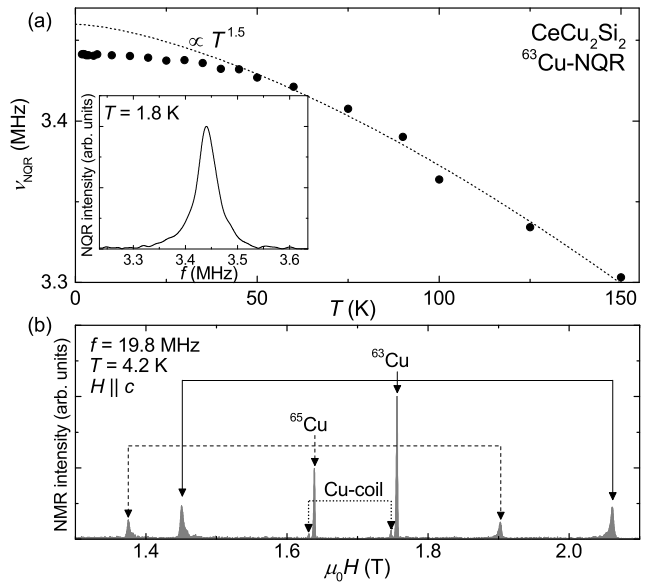}
\end{center}
\caption{(a)Temperature dependence of $^{63}$Cu-NQR frequency.
The dotted line is an empirical relation of $\nu_Q(T) =\nu_Q(0) (1-\alpha T^{3/2})$.
(Inset) Frequency dependence of $^{63}$Cu-NQR spectrum at 1.8~K. 
(b) Field-swept NMR spectrum at 4.2~K and $f$ = 19.8~MHz for $H \parallel c$.
}
\label{Fig.1}
\end{figure}
The inset of Fig.~\ref{Fig.1}(a) shows the $^{63}$Cu-NQR spectrum as a function of frequency.
When $I \ge 1$, the nucleus has an electric quadrupole moment $Q$ as well as a magnetic dipole moment; thus, the degeneracy of the nuclear-energy levels is lifted even at zero magnetic field due to the interaction between $Q$ and the electric field gradient (EFG) $V_{zz} = eq$ at the nuclear site.
The electric quadrupole Hamiltonian $\mathcal{H}_Q$ can be described as
\begin{align}
\mathcal{H}_Q &= \frac{\nu_{zz}}{6}\left\{(3I_z^2-I^2)+\frac{1}{2}\eta(I_+^2+I_-^2)\right\},
\label{eq.1}
\end{align}
where $\nu_{zz}$ is the quadrupole frequency along the principal axis ($c$-axis) of the EFG, defined as $\nu_{\rm zz} \equiv 3e^2qQ/2I(2I-1)$ with $eq = V_{zz}$,  and $\eta$ is the asymmetry parameter of the EFG expressed as $(V_{xx}-V_{yy})/ V_{zz}$ with $V_{\alpha \alpha}$, which is the second derivative of the electric potential $V$ along the $\alpha$ direction ($\alpha = x,y,z$).
The parameter $\eta$ should be zero at the Cu site in CeCu$_2$Si$_2$ because of the four-fold symmetry.
The obtained NQR frequency $\nu_{\rm NQR}$ = 3.441~MHz at 1.8~K was almost the same as that in the polycrystalline samples.
The full width at half maximum (FWHM) in the $^{63}$Cu-NQR spectrum, which depended on crystalline homogeneity, was 41~kHz and was almost temperature independent.
The obtained FWHM was broader than that in high-quality polycrystalline CeCu$_{2.05}$Si$_2$ (FWHM $\sim$ 13~kHz) characterized as an $A/S$-type sample and that in Ce$_{1.025}$Cu$_{2}$Si$_2$ (FWHM $\sim$ 26~kHz) characterized as an $S$-type sample.
The FWHM result indicated that the crystal homogeneity in the present single-crystal sample was not as good as that in the polycrystalline $A/S$-type CeCu$_{2.05}$Si$_2$.
This is consistent with previous results that an $S$-type sample is located at the Cu-rich region in the qualitative Ce--Cu--Si phase diagram of CeCu$_2$Si$_2$\cite{F.Steglich_PhysicaB_1996}.     

As shown in Fig.~\ref{Fig.1}(a), $\nu_{\rm NQR}$ increases with decreasing temperature.
The temperature variation of $\nu_{\rm NQR}$ followed the empirical relation of $\nu_Q(T) =\nu_Q(0) (1-\alpha T^{3/2})$ down to 50~K owing to a thermal lattice expansion and/or lattice vibrations\cite{J.Christiansen_ZPB_1976,H.Nakamura_JPSJ_1990,S.-H.Baek_PRB_2009_2} and deviated downward from the relation. 
Similar temperature dependence has been observed in various Ce-based filled skutterudites\cite{K.Magishi_JPSJ_2012,M.Yogi_JPSCP_2013}.
No clear change of $\nu_Q$ was observed around 15 K, where the 4$f$ electron character changed from a localized to itinerant nature, as we discuss later. 
This suggested that the Ce valence in CeCu$_2$Si$_2$ did not change when the HF state was formed at ambient pressure. 

\begin{figure}[!tb]
\vspace*{10pt}
\begin{center}
\includegraphics[width=8cm,clip]{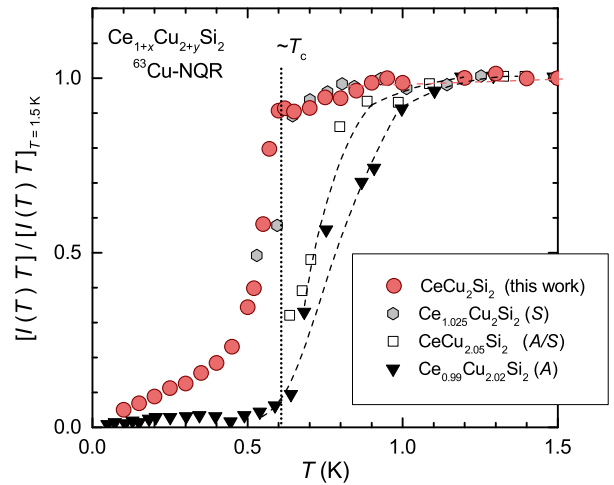}
\end{center}
\caption{(Color online) Temperature dependence of the Cu-NQR intensity ($I$) multiplied by $T$, $I(T) T$, normalized by $IT$ at 1.5~K for the present single-crystal CeCu$_2$Si$_2$, and compared with the various polycrystalline samples\cite{K.Ishida_PRL_1999}.
The dotted line indicates $T_{\rm c}$, and the broken lines provide a guide to the eye.}
\label{Fig.2}
\end{figure}

Figure \ref{Fig.2} shows the temperature dependence of the $^{63}$Cu-NQR intensity ($I$) multiplied by $T$, $I(T) T$, which is normalized by $IT$ at 1.5~K for the present single-crystal CeCu$_2$Si$_2$, compared to various polycrystalline samples\cite{K.Ishida_PRL_1999}. 
The value of $IT$ decreases rapidly below $T_{\rm c}$ due to the SC shielding effect of the rf field.
As we reported in previous papers\cite{K.Ishida_PRL_1999}, $IT$ in the $A$ and $A/S$-type samples decreased significantly below about 1.0~K due to the appearance of the magnetic fraction related to the $A$ phase.  
On the other hand, the loss of the NQR intensity in the $S$-type polycrystalline Ce$_{1.025}$Cu$_2$Si$_2$ was small down to $T_{\rm c}$. 
Since the temperature dependence of $IT$ in the present single-crystal CeCu$_2$Si$_2$ was similar to that of the $S$-type polycrystalline Ce$_{1.025}$Cu$_2$Si$_2$, the present single crystal was also characterized as an $S$-type sample. 
       
\begin{figure}[!b]
\vspace*{10pt}
\begin{center}
\includegraphics[width=8cm,clip]{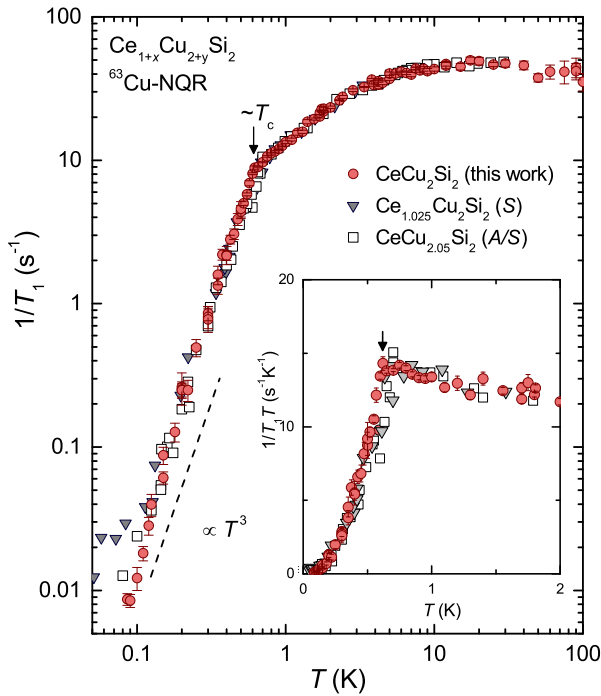}
\end{center}
\caption{(Color online) Temperature dependence of $1/T_1$ measured with NQR on the present $S$-type single-crystal CeCu$_2$Si$_2$. 
The NQR-$1/T_1$ results on the polycrystalline $S$-type Ce$_{1.025}$Cu$_{2}$Si$_2$ and $A/S$-type CeCu$_{2.05}$Si$_2$ are also plotted\cite{K.Ishida_PRL_1999}.
The linear scale plot of $1/T_1T$ around $T_{\rm c}$ is shown in the inset.}
\label{Fig.3}
\end{figure}

Figure \ref{Fig.3} shows the temperature dependence of $1/T_1$ of the single-crystal CeCu$_2$Si$_2$, along with those of the polycrystalline $S$-type Ce$_{1.025}$Cu$_{2}$Si$_2$ and $A/S$-type CeCu$_{2.05}$Si$_2$, measured by $^{63}$Cu-NQR.
In the present single crystal, $1/T_1$ was quite similar to $1/T_1$ in the polycrystalline samples. In all samples, $1/T_1$ was almost constant at high temperatures and started to decrease below $T^* \sim$ 15~K. 
Here, $T^*$ is defined as the characteristic temperature of the Ce - 4$f$ electrons.
With further cooling, 
$1/T_1T$ in the single-crystal sample showed almost constant behavior below 0.8~K. 
The formation of the Fermi-liquid state above $T_{\rm c}$ is one of the characteristic features of $S$-type samples.
On the other hand, the $A/S$-type sample showed that $1/T_1T$ continued to increase down to $T_{\rm c}$ accompanied with the gradual decrease of the NQR signal intensity. 
These are the anomalies related to the $A$-phase.   

In the SC state, 1/$T_1$ in all samples showed no clear coherence (Hebel-Slichter) peak just below $T_{\rm c}$, and 1/$T_1$ was proportional to $T^{3}$ at low temperatures down to 130~mK.
The $T^{3}$-dependence of 1/$T_1$ was consistent with the $T$-linear dependence of $C/T$ in the intermediate temperature range between $T_{\rm c}$ and 200~mK. 
Below 120~mK, 1/$T_1$ in the single-crystal sample deviated downward from the $T^{3}$-dependence, which was consistent with the exponential behavior of $C/T$ in the temperature region between 50 and 200~mK\cite{S.Kittaka_PRL_2014}.
Low-temperature 1/$T_1$ below 90~mK could not be measured due to the limits of the refrigerator in our lab.
Possible gap structure will be discussed based on the temperature dependence of $1/T_1$ in the single-crystal sample later in the discussion part.  

For the NMR measurement, we applied magnetic fields to lift the degeneracy of the spin degrees of freedom, even though the nuclear-energy levels were already split by the electric quadrupole interaction.
Total effective Hamiltonian could be expressed as
\begin{align}
\mathcal{H} &= \mathcal{H}_{\rm Z} + \mathcal{H}_{\rm Q} \notag \\
            &= -\gamma \hslash (1 + K)I \cdot H + \mathcal{H}_{\rm Q},
\end{align}
where $K$ is the Knight shift, and $H$ is an external field.
Four nuclear spin levels were well separated, and we observed three resonance lines for each isotope ($^{63}$Cu and $^{65}$Cu) as shown in Fig.~\ref{Fig.1}(b).
Since the position of the resonance line depended on the angle between the applied magnetic field and the principal axis of the EFG ($\parallel$ $c$ axis in CeCu$_2$Si$_2$), we could determine the field direction with respect to the $c$-axis from the NMR peak locus.
The misalignment of the $c$-axis with respect to the field-rotation plane was estimated to be less than 2$^{\rm o}$ from the NMR spectrum analyses, and
$K$ was determined from the central line of the $^{63}$Cu-NMR spectrum.

\begin{figure}[!tb]
\vspace*{10pt}
\begin{center}
\includegraphics[width=8cm,clip]{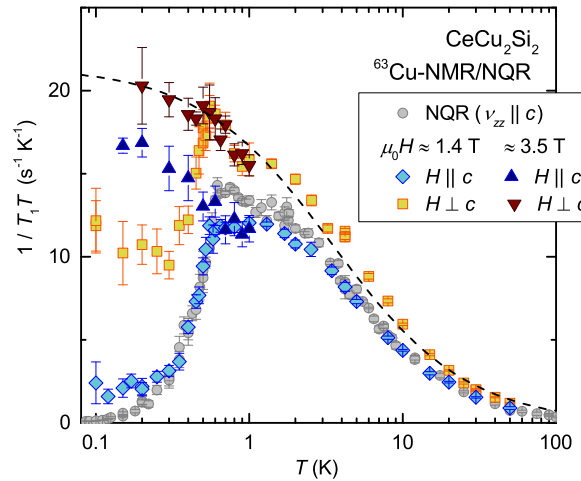}
\end{center}
\caption{(Color online)Temperature dependence of $1/T_1T$ on the present single crystal at 0~T (NQR), 1.4~T, and 3.5~T for $H \parallel$ c and $H \perp c$.
The dotted line is a Curie--Weiss dependence estimated from the fitting below 2~K [$C/(T+\theta)$ with $C$ = 75 s$^{-1}$ and $\theta$ = 3.5 K]. 
The small $\theta$ indicates that the system is close to a quantum critical point.}
\label{Fig.4}
\end{figure}
Figure \ref{Fig.4} shows the temperature dependence of $1/T_1T$ at zero field, 1.4 T ($< \mu_0 H_{\rm c2}$), and 3.5 T ($> \mu_0 H_{\rm c2}$) parallel and perpendicular to the $c$ axis.  
In the normal state, $(1/T_1T)_{H \perp c}$ was larger than $(1/T_1T)_{H \parallel c}$ by a factor of 1.32 [$(1/T_1T)_{H \perp c} = 1.32 (1/T_1T)_{H \parallel c}$], while the temperature dependence was almost identical between the two directions.
The anisotropy of $1/T_1T$ was considered to originate from the anisotropy of the hyperfine coupling constant and spin susceptibility.
As mentioned above, $1/T_1T$ measured at zero field became constant below 0.8~K, but $1/T_1T$ continued to increase as the temperature decreased to 150~mK when superconductivity was suppressed by the field above $\mu_0 H_{\rm c2}$. 
In field lower than $\mu_0 H_{\rm c2}$, constant $1/T_1T$ was observed at low temperatures in the SC state, which was indicative of the presence of the field-induced residual density of states ascribed to vortex cores.

\begin{figure}[!tb]
\vspace*{10pt}
\begin{center}
\includegraphics[width=8cm,clip]{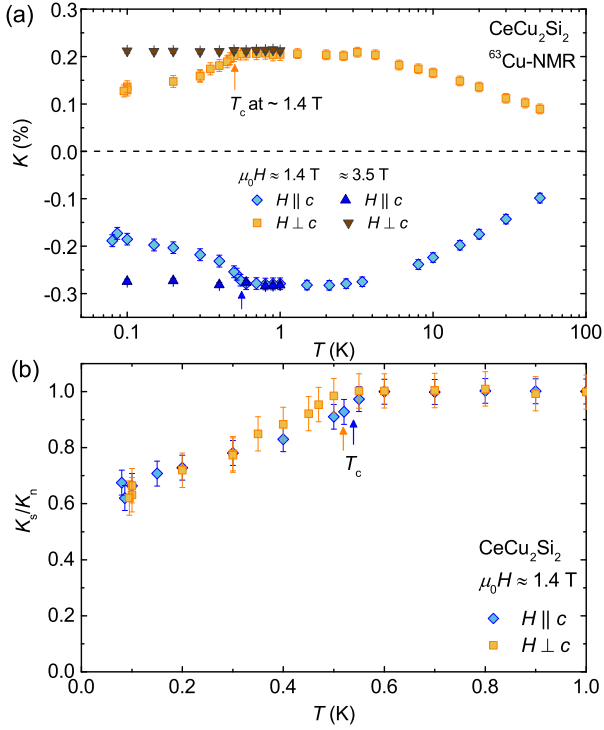}
\end{center}
\caption{(Color online)(a) Temperature dependence of the Knight shift at 1.4~T and 3.5~T for $H\parallel c$ and $H\perp c$.
In contrast with constant behavior below 1~K at 3.5~T ($> \mu_0H_{\rm c2}$), the absolute value of $K_{i}$ decreases below $T_{\rm c}$ at 1.4~T, reflecting the decrease of the spin susceptibility in the SC state.
(b)Temperature dependence of spin susceptibility normalized at $T_{\rm c}$.
}
\label{Fig.5}
\end{figure}

Figure~\ref{Fig.5}(a) shows the temperature dependence of $K_{i}$ ($i = \perp$ and $c$) measured at 1.4~T and 3.5~T for both directions.
The Knight shift $K_{i}$ is described as
\begin{align}
K_{i} = A_{{\rm hf},i} \chi_{{\rm spin},i} + K_{{\rm orb},i},
\label{eq.K}
\end{align}
where $A_{{\rm hf},i}$, $\chi_{{\rm spin},i}$, and $K_{{\rm orb},i}$ are the hyperfine coupling constant, spin susceptibility, and orbital part of the Knight shift in each direction, and $K_{{\rm orb},i}$ is usually temperature independent.
In the normal state, $K_{\perp}$ increased on cooling and became constant below 4~K.
The temperature dependence of $K_{c}$ was similar to that of $K_{\perp}$, with opposing sign due to the anisotropic $A_{\rm hf}$, which is understood by c-f hybridization.\cite{T.Ohama_JPSJ_1995}
In contrast to the constant behavior below 1~K in 3.5~T ($> \mu_{0}H_{\rm c2}$), the absolute value of $K_{i}$ decreased below $T_{\rm c}$ at 1.4~T, indicative of the decrease of the spin susceptibility in the SC state. 
This decrease will be discussed quantitatively later.  

\section{Discussion}
\subsection{Spin dynamics in the normal state}
In general, $1/T_1$ provides microscopic details about the low-energy spin dynamics, and thus, we analyze $1/T_1$ to quantitatively discuss the character of low-energy spin dynamics of Ce moments.
In temperatures higher than the coherent temperature $T^*$, the Ce moments are in a well localized regime; thus, the observed $1/T_1$ value in CeCu$_2$Si$_2$ is approximately decomposed into conduction-electron and localized Ce $f$-electrons as 
\begin{equation}
(1/T_1)_{\rm obs} = (1/T_1)_c + (1/T_1)_f, 
\end{equation}
where the former contribution can be approximately known from $1/T_1$ of the LaCu$_2$Si$_2$\cite{T.Ohama_PhDthesis_1995}.
The latter contribution is dominated by fluctuations of the Ce spins and can be given by the Fourier component of $\langle S(t)S(0) \rangle$ at the Larmor frequency, 
where the time dependence arises from the fluctuations of the Ce spins.

In general, $1/T_1$ is expressed as\cite{T.Moriya_PTP_1962}
\begin{equation}
\frac{1}{T_1} = \frac{\gamma_n^2k_{\rm B}T}{2\mu_{\rm B}^2}\lim_{\omega \rightarrow 0} \sum_{q} [A(q)]^2 \frac{\chi''(q,\omega)}{\omega},
\label{eq.5}
\end{equation}
where $A(q)$ is the $q$-dependent hyperfine coupling constant, and $\chi''(q,\omega)$ is the imaginary part of the dynamical susceptibility, and the sum is over the Brillouin zone.
At higher temperatures, the spin dynamics are determined by independent Ce moments, and the local-moment susceptibility is given by\cite{D.L.Cox_JAP_1985}
\begin{equation}
\chi_L(\omega) =\frac{\chi_0(T)}{1- i \omega / \Gamma(T)},
\end{equation}
where $\chi_0$ is the bulk susceptibility and $\Gamma$ is the characteristic energy of spin fluctuations of Ce moments.

We assume that the $q$-dependence of $A(q)$ can be negligibly small, and the dynamical susceptibility is isotropic.
Then, eq.\eqref{eq.5} can be described as\cite{D.E.MacLaughlin_JAP_1979,D.E.MacLaughlin_PRB_1981}
\begin{equation*}
\left(\frac{1}{T_1}\right)_f \sim \frac{N \gamma_n^2 k_{\rm B}T A^2}{\mu_{\rm B}^2}\frac{\pi \hbar \chi_0(T)}{\Gamma(T)}, 
\end{equation*}
where $(1/T_1)_f$ is estimated by subtracting $1/T_1$ of LaCu$_2$Si$_2$ from $1/T_1$ of CeCu$_2$Si$_2$ measured with the $^{63}$Cu-NQR, and $N$ is the number of the nearest neighbor Ce sites.  
Using this equation, $\Gamma(T) / k_{\rm B}$ is expressed with the NMR quantities as 
\begin{equation}
\frac{\Gamma(T)}{k_B} = N \gamma_n^2\pi \hbar \left(\frac{A_{\perp}}{\mu_B}\right)T K_{\perp} (T_1)_f,
\label{eq.8}
\end{equation}
where $K_{\perp}$ is the Cu Knight shift perpendicular to the $c$ axis. Here, $A_{\perp}$ is the hyperfine coupling constant perpendicular to the $c$ axis, which is evaluated from the $K$-$\chi$ plot in the $T$ range from 8 and 80~K\cite{T.Ohama_JPSJ_1995}, since the bulk susceptibility is easily affected by an extrinsic impurity contribution.
     
\begin{figure}[tbp]
\begin{center}
\includegraphics[width=8cm,clip]{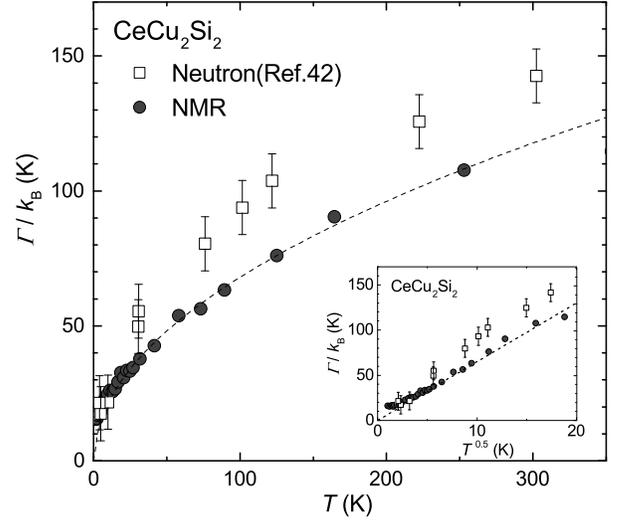}
\end{center}
\caption{Temperature dependence of the characteristic energy of the spin fluctuations $\Gamma(T)$ evaluated with the NMR quantities is shown, along with the temperature dependence of the half-width of the quasielastic neutron scattering line.     
The dotted curve is the $T^{1/2}$ dependence, which is a high-temperature approximation of the theoretical calculation of $\Gamma$ based on the impurity Kondo model by Cox {\it et al.}\cite{D.L.Cox_JAP_1985}. The fitting is fairly good above 20 K. 
(Inset) Temperature dependence of the characteristic energy of the spin fluctuations $\Gamma(T)$ as a function of the square root of T.}
\label{Fig.6}
\end{figure}
Figure \ref{Fig.6} shows the temperature dependence of $\Gamma(T)/k_{\rm B}$ estimated by eq.\eqref{eq.8}, as well as $\Gamma(T)/k_{\rm B}$ directly measured with neutron quasielastic scattering (NQS)\cite{S.Horn_PRB_1981}.
A similar comparison has been performed with $^{29}$Si-NMR results on a polycrystalline CeCu$_2$Si$_2$\cite{J.Aarts_Physica_1983}, but the agreement was not as good as that from the current study, probably due to the impurity-phase contribution in the bulk susceptibility.   
In the present analyses based on the $^{63}$Cu-NMR results, the agreement is rather good, and both $\Gamma(T)/k_{\rm B}$ show very similar $T$ dependence, although the NQS result is somewhat larger than the NMR estimation.  
In particular, $\Gamma(T)/k_{\rm B}$ follows a $T^{1/2}$ dependence above 20~K.
In HF compounds containing Ce and Yb ions, $\Gamma(T)$ was calculated for independently screened local moments based on an impurity-Kondo model for Ce$^{3+}$($4f^1$) and Yb$^{3+}$($4f^{13}$) by Cox {\it et al.}\cite{D.L.Cox_JAP_1985}
The $T^{1/2}$ dependence is the high-temperature approximation of the theoretical calculation of $\Gamma/k_{\rm B}$ and has been observed in various HF compounds.
As shown in Fig.~\ref{Fig.6}, $\Gamma/k_{\rm B}$ deviated from the $T^{1/2}$ dependence and remained at constant value below around 15~K owing to the formation of the low-temperature coherence ground state.
In fact, the resistivity showed broad maximum at around 15~K, and thus, the resistivity and $1/T_1$ results showed the occurrence of local-moment screening below 15~K by the ``{\it Kondo effect}".

As shown in Fig.~\ref{Fig.5}(a), the static susceptibility became constant below 4~K, whereas $1/T_1T$ probing $q$-summed dynamical susceptibility continued to increase as temperature decreased to 0.8~K at zero field. 
Thus, AFM fluctuations become dominant at low temperatures.
The nature of the AFM fluctuations was investigated by neutron scattering measurements and is revealed to be of the incommensurate SDW-type with a propagation vector $\mbox{\boldmath $Q$}_{\rm AF}$ = (0.22, 0.22, 0.53), which is the same propagation vector of the $A$-phase ordered state\cite{O.Stockert_PhysicaB_2008,O.Stockert_NaturePhys_2011}.

Finally, we discuss the possibility of the field-induced AFM state in the present $S$-type CeCu$_2$Si$_2$.
The field-induced magnetic anomaly was reported from magnetoresistance and de Haas--van Alphen measurements in a previous single-crystal sample\cite{F.Steglich_JPCM_1989,M.Hunt_JPCM_1990}.
In general, when magnetic ordering occurs, $1/T_1T$ shows a peak at magnetic ordering temperature $T_{\rm M}$, and the NMR spectra show broadening and/or splitting below $T_{\rm M}$.
However, in this study, $1/T_1T$ does not show such a peak but continues to increase as the temperature decreases to 150~mK, following the Curie--Weiss dependence shown by the dotted curve in Fig.~\ref{Fig.4} when 3.5~T ( $> \mu_0H_{\rm c2}$) is applied perpendicularly to the $c$ axis.
A similar continuous increase of $1/T_1T$ was observed in the field parallel to the $c$ axis, indicating the development of AFM fluctuations.
The small but finite Weiss temperature estimated from the fitting below 2~K ($\theta \sim 3.5$~K) suggests that the present $S$-type sample is still in the paramagnetic state, although it is close to a quantum critical point. 
These results are consistent with recent neutron scattering results\cite{J.Arndt_PRL_2011}.
In addition, no clear reduction of NMR intensity related to the $A$-phase anomaly was observed\cite{H.Nakamura_JMMM_1988}. 
Our NMR results indicate the absence of the field-induced magnetic anomaly in the present $S$-type single crystal.

\subsection{Superconducting gap symmetry}
\begin{figure}[!tb]
\vspace*{10pt}
\begin{center}
\includegraphics[width=7cm,clip]{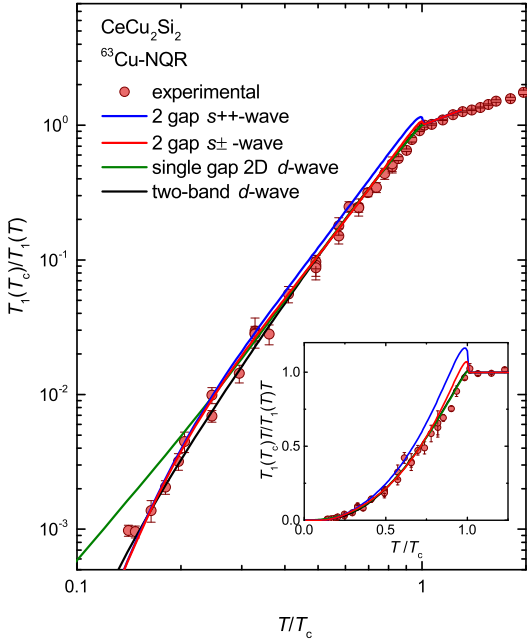}
\end{center}
\caption{(Color online) Log-log plot of the calculations of normalized $1/T_1$ with each SC model, and the experimental result of the normalized $1/T_1$ results at zero field. 
Inset shows the linear scale plot of normalized $1/T_1T$ and the calculations.}
\label{Fig.7}
\end{figure}

\begin{figure}[!tb]
\vspace*{10pt}
\begin{center}
\includegraphics[width=8cm,clip]{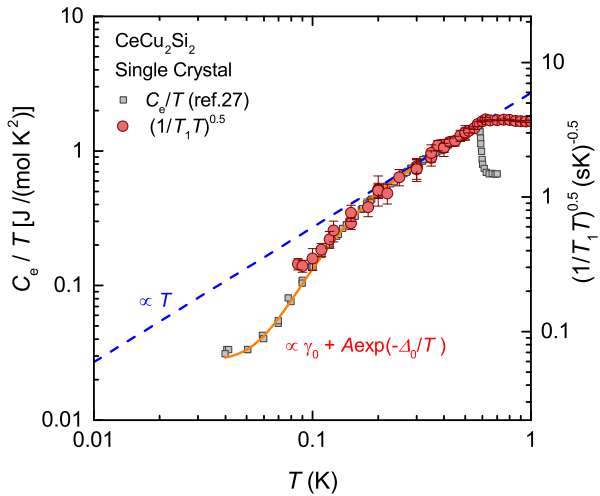}
\end{center}
\caption{(Color online) Log-log plot of the specific heat $C$ divided by temperature\cite{S.Kittaka_PRL_2014} and the square root of $1/T_1T$ of $S$-type CeCu$_2$Si$_2$.
The broken and dotted lines are plotted to guide the eye.}
\label{Fig.8}
\end{figure}

Here, we discuss a plausible SC gap model for explaining the temperature variation of $1/T_1$ at zero field.
The $1/T_1$ results showing $T^{3}$ dependence were considered to be evidence of the presence of a line node in CeCu$_2$Si$_2$, and these results can be reproduced by the two-dimensional $d$-wave model, as shown in Fig.~\ref{Fig.7}.
However, recent specific heat measurements indicate the absence of nodal quasi-particle excitations and the presence of a finite gap with a small magnitude of $\Delta_0 \sim$ 0.30~K ($\sim 0.43 T_{\rm c}$) at low temperatures, although $C/T$ increases linearly with temperature for $T > 0.2$~K as shown in Fig.~\ref{Fig.8}.
These results, as well as the absence of $C/T$ oscillation in the field-angle dependence measurements, suggest that CeCu$_2$Si$_2$ is a multiband full-gap superconductor.
In addition, a multiband full-gap superconductor without sign change ($s_{++}$-wave) and a fully gapped two band $d$-wave superconductor (two-band $d$-wave) were recently proposed by electron irradiation experiments\cite{T.Yamashita_SciAdv_2017} and penetration depth measurements\cite{G.M.Pang_arXiv_2016}, respectively.
A multigap SC model with more than two full gaps of different gap sizes was not general before the discovery of Sr$_{2}$RuO$_{4}$\cite{Y.Maeno_Nature_1994,Y.Maeno_JPSJ_2012}, MgB$_2$\cite{J.Nagamatsu_Nature_2001,C.Buzea_SST_2001}, and Fe-based superconductors\cite{Y.Kamihara_JACS_2008,K.Ishida_JPSJ_2009,J.Paglione_Naturephys_2010}, and thus, such a multigap model was not applied to reproduce experimental results in unconventional superconductors before 2000.
Furthermore, owing to the complex Fermi surfaces in HF superconductors, the single-band analysis was conventionally adopted for simplicity.
However, after the discovery of the Fe-based superconductors, it was clear that the $T^{3}$ dependence of $1/T_1$ could be reproduced not only by the line nodal SC gap but also by the multiband full-gap.
In fact, the low-temperature $T^3$ behavior of $1/T_1$ observed in LaFeAs(O$_{0.89}$F$_{0.11}$) is not consistent with the $d$-wave model with line nodes since deviation of the $T^3$ dependence, which is expected in a $d$-wave superconductor, was not observed even in inhomogeneous samples, as shown with $^{75}$As-NQR measurements\cite{Y.Nakai_JPSJ_2008,S.Kitagawa_PhysicaC_2010}.
Furthermore, the multiband full-gap structure was actually detected from angle-resolved photoemission spectroscopy\cite{H.Ding_EPL_2008}, and thus, the multiband SC model has been accepted as a realistic model for interpreting experimental results.
Therefore, as already discussed by Kittaka {\it et al.}\cite{S.Kittaka_PRL_2014}, we must identify whether the present NQR results can be consistently understood by the two-band SC model.

The temperature dependence of 1/$T_1T$ in two-gap superconductors is calculated using the following equations:
\begin{align}
\frac{1}{T_1T} \propto &\int^{\infty}_{0} \left\{\left[\sum_{i} N_{s}^{i}(E)\right]^{2} + \left[\sum_{i} M_{s}^{i}(E)\right]^{2} \right\} \notag 
\\ 
&\times f(E)[1-f(E)]{\rm d}E,\notag \\
N_{s}^{i}(E) = &n_{i} \int^{\infty}_{0} \frac{E'}{\sqrt{E'^2 - \Delta_{i}^2}} \frac{1}{\sqrt{2\pi \delta_{i}^2}} \exp\left[-\frac{(E-E')^2}{2\delta_{i}^2}\right] {\rm d}E',\notag \\
M_{s}^{i}(E) = &n_{i} \int^{\infty}_{0} \frac{\Delta_{i}}{\sqrt{E'^2 - \Delta_{i}^2}} \frac{1}{\sqrt{2\pi \delta_{i}^2}} \exp\left[-\frac{(E-E')^2}{2\delta_{i}^2}\right] {\rm d}E'. \notag 
\end{align}
Here, $N_{s}^{i}(E), M_{s}^{i}(E)$, $\Delta_{i}$, $\delta_{i}$, and $f(E)$ are the quasi-particle density of state (DOS), the anomalous DOS arising from the coherence effect of Cooper pairs, the amplitude of the SC gap, the smearing factor to remove divergence of $N_{s}^{i}(E)$ at $E = \Delta_{i}$, and the Fermi distribution function, respectively.
The parameter $n_i$ represents the fraction of the DOS of the $i$-th SC gap, and two SC gaps are assumed for simplicity, thus $n_1 + n_2 = 1$. 
We multiply $N_{s}^{i}(E)$ and $M_{s}^{i}(E)$ by a Gaussian distribution function to suppress the coherence peak.
We also calculate 1/$T_1T$ using a single-gap two dimensional $d$-wave model and a two-band $d$-wave model discussed in ref.~46 as follows:
\begin{align}
\frac{1}{T_1T} &\propto \int^{\infty}_{0} N_{s}^{d}(E)^{2} f(E)[1-f(E)]{\rm d}E, \notag \\
N_{s}^{d}(E) &= \int^{2\pi}_{0}\frac{{\rm d}\phi}{4\pi} \int^{\pi}_{0} {\rm d}\theta \sin\theta \frac{E}{\sqrt{E^2 - \Delta(\theta,\phi)^2}}, \notag \\
\Delta(\theta,\phi) &= \Delta_0\cos(2\phi) \text{\hspace*{10pt} (single-gap $d$-wave)}, \notag \\
\Delta(\theta,\phi) &= \sqrt{[\Delta_1\cos(2\phi)]^2 + [\Delta_2\sin(2\phi)]^2} \notag \\
&\text{\hspace*{100pt} (two-band $d$-wave)}, \notag
\end{align}
where $N_{s}^{d}(E)$ is the quasi-particle DOS in a $d$-wave superconductor and $\Delta_0$ is the maximum of the SC gap.

Figure~\ref{Fig.7} shows the calculated $1/T_1$ in each model together with experimental data as a function of the normalized temperature.
All parameters used for the calculations are listed in Table~\ref{Tab.1}.
The $1/T_1T$ behavior in the two-gap $s_{++}$-wave shows a clear coherence peak, which seems to be inconsistent with the experimental results.
As discussed by Kittaka {\it et al.}\cite{S.Kittaka_PRB_2016}, large and/or temperature-dependent smearing factors originating from quasiparticle damping by AFM fluctuations might suppress the coherence peak. 
However, such a large smearing factor generally suppresses the SC transition temperature.
In addition, the coherence peak was not observed even in pressure-applied CeCu$_2$Si$_2$, where the AFM fluctuations were significantly suppressed\cite{K.Fujiwara_JPSJ_2008}.
Thus, the suppression of the coherence peak by the damping effect of AFM fluctuations seems to be unlikely.        
Rather, the two-gap $s_{\pm}$-wave, two-dimensional $d$-wave, and two-band $d$-wave can closely reproduce the experimental results near $T_{\rm c}$.
The experimental $1/T_1$ value deviated from $T^{3}$ behavior below 0.2 $T_{\rm c}$, which agreed with the two-gap $s_{\pm}$-wave and two-band $d$-wave behavior.
However, the $d$-waves seem inconsistent with the absence of the oscillation of $C/T$ in the field-angle dependence\cite{S.Kittaka_PRL_2014}.
We can safely say that $1/T_1T$ results down to 90~mK  can be reproduced by the two-gap $s_{\pm}$-wave, which was suggested by recent specific heat measurements\cite{S.Kittaka_PRL_2014}.
In fact, the square root of $1/T_1T$ shows almost the same temperature dependence as $C_{\rm e}/T$ down to 90~mK, as shown in Fig.~\ref{Fig.8}.

In the plausible $s_{\pm}$ state of CeCu$_2$Si$_2$, the sign of the SC gap would change at the electron Fermi surface that is located around the X point with a loop-shaped node.
However, as suggested by Ikeda {\it et al.}, because this nodal feature is not symmetry protected, the loop node can be easily lifted by the slight mixture of on-site pairing due to an intrinsic attractive on-site interaction, and the corrugated heavy-electron sheet becomes fully gapped with a small magnitude of the SC gap\cite{H.Ikeda_PRL_2015}. 
The small full gap observed by various experiments in CeCu$_2$S$_2$ can be understood by this scenario.

Recently, Yamashita {\it et al.}\cite{T.Yamashita_SciAdv_2017} reported that the superconductivity of {\it S}-type CeCu$_2$Si$_2$ is robust against the impurity scattering induced by electron-irradiation-creating point defects, which strongly suggested that the superconductivity is of the $s_{++}$-wave type without sign reversal. 
As mentioned above, the $s_{++}$-wave seems to be inconsistent with the temperature dependence of $1/T_1$ just below $T_c$.
The absence of the coherence peak immediately below $T_{\rm c}$ and the robustness of superconductivity against the impurity scattering should be interpreted on the same footing. 
The same discrepancy has been also identified in an iron-based superconductor with the ``1111'' structure\cite{M.Sato_JPSJ_2010}.      
To settle this discrepancy, the Fermi-surface properties of CeCu$_2$Si$_2$ should be clarified with experiments such as de Haas--van Alphen, angle-resolved photo-emission spectroscopy, and scanning tunneling microscope measurements.     
      
\begin{table}[bp]
\caption[]{Superconducting gaps $\Delta_i$, smearing factor $\delta_i$, and weight of the primary band used for the calculation of $T_1$.} 
\label{table:Sb_parameter}
\vspace{1cm}
\begin{tabular}{cccccc}
\hline
Model & $\Delta_{1}$ & $\Delta_{2}$ & $\delta_{1}$/$\Delta_{1}$  &  $\delta_{2}$/$\Delta_{2}$ &$ n_1$ \\ \hline
2-gap $s_{++}$ & 2.1 & 0.8 & 0.2 & 0.2 & 0.65 \\
2-gap $s_{\pm}$ & 2.1 & -0.8 & 0.2 & 0.2 & 0.65  \\
1-gap $d$ & 2.1 & - & - & - & 1.0  \\
two-band $d$ & 2.1 & 0.4 & - & - & 1.0  \\
\hline
\end{tabular}
\label{Tab.1}
\end{table}

Finally, we illustrate the differences between $1/T_1$ of CeCu$_2$Si$_2$ and $1/T_1$ of CeCoIn$_5$ in the SC state.
Various experiments have suggested the presence of a line node in CeCoIn$_5$ not only from the temperature dependence but also from the field-angle dependence, and CeCoIn$_5$ is considered to be of $d$-wave symmetry\cite{K.Izawa_PRL_2001,K.An_PRL_2010,Y.Kohori_PRB_2001}.
Although both compounds show similar temperature dependence of $1/T_1$ ($1/T_1 \propto T^3$) and the absence of a coherence peak immediately below $T_{\rm c}$, a clear difference was observed at low temperatures.
As shown in Fig. \ref{Fig.3}, $1/T_1$ shows  a $T^3$ dependence down to 130~mK, but $1/T_1$ of CeCoIn$_5$ deviated {\it upward} from the $T^3$ dependence below 300~mK and showed $T$-linear behavior below 100~mK\cite{Y.Kohori_PRB_2001,Y.Kawasaki_JPSJ_2003}.
The deviation seems to depend on the quality of the samples: larger deviations are observed in lower quality samples.
Because this deviation, which originates from the residual DOS at the Fermi energy, has been commonly observed in unconventional superconductors with  symmetry-protected line nodes such as cuprate superconductors\cite{K.Ishida_PhysicaC_1991,K.Ishida_JPSJ_1993}, the absence of an appreciable deviation from the $T^3$ dependence even in non-stoichiometric CeCu$_2$Si$_2$ cannot be understood by such a line node.
Instead, this result does suggest that the SC state is not a $d$-wave.

\subsection{Spin susceptibility below $T_{\rm c}$}

Next, we discuss the spin susceptibility in the SC state.
The Knight shift measurement in the SC state is known to be one of few measurements to give information about the spin state of superconductors. 
Since the Knight shift consists of spin and orbital components, as shown in eq.~\eqref{eq.K}, we need to estimate the orbital part to determine the spin susceptibility.
Ohama {\it et al.} measured the Knight shift and $1/T_1T$ of $^{29}$Si and $^{63}$Cu in a magnetically-aligned powder sample of CeCu$_{2}$Si$_2$ and reported that the Knight shift and $1/T_1T$ of the Cu site were determined by a conduction-electron effect at higher temperature regions. The present $1/T_1T$ value and Knight shift at high temperatures in CeCu$_{2}$Si$_2$ were similar values as YCu$_2$Si$_2$\cite{T.Ohama_JPSJ_1995}.
Thus, we assume $K_{\rm orb} \sim 0 $ at both directions, as in the case of YCu$_2$Si$_2$.
Figure~\ref{Fig.5}(b) shows the temperature dependence of the spin component of the Knight shift ($K_s$) normalized by the value at $T_{\rm c}$ ($K_n$).
Here, $(K_s/K_n)_{H \parallel c}$ = $(K_s/K_n)_{H \perp c}$ = 0.6 at the lowest temperature under $\mu_0 H \sim$ 1.4 T.
This residual Knight shift originated from the field-induced normal state owing to vortex cores because $K_s/K_n$ at the lowest temperature became smaller in lower fields and thus the spin susceptibility would become zero at 0~K near zero fields, which provides strong evidence of a spin-singlet superconductor\cite{Y.Kitaoka_JJAP_1987}.
However, the residual normalized DOS estimated from $1/T_1T$ was 0.4 for $H \parallel c$ and 0.7 for $H \perp c$, which was slightly different from the estimation from $K_s/K_n$.
We propose this discrepancy to be due to the SC diamagnetic field. 
Assuming the residual $K_s/K_n$ to be equal to the residual DOS (estimated from $1/T_1T$) implies a diamagnetic Knight shift $K_{\rm dia}$ of about 0.03~\%.
In fact, $K_{\rm dia}$ is estimated as 0.03~\% from the formula of $H_{\rm dia} = H_{c1} [\ln(\beta d/\sqrt(e))/\ln(\kappa)]$.
Here, the lower critical field $H_{c1} = 30$~Oe, $\beta = 0.38$ in the triangular vortex lattice, the distance between vortices $d = 412$~\AA~at 1.4~T, and the Ginzburg--Landau parameter $\kappa = 141$ are used for the estimation\cite{S.Kittaka_PRB_2016,A.Pollini_JLTP_1993}.
These results suggest that the spin susceptibility in both directions becomes zero near zero field in CeCu$_{2}$Si$_{2}$ because $1/T_1T$ at the lowest temperatures becomes zero at low fields.
Note that the normal-state $K_s$, which was enhanced with decreasing temperature, disappeared completely below $T_{\rm c}$ in CeCu$_2$Si$_2$, which is indicative of singlet pairing by the pseudo-spin $J$.
On the other hand, the decrease of $K_s$ in the SC state is usually very small in U-based heavy-fermion superconductors.
In addition, even in Ce compounds, the decrease of $K_s$ is small in noncentrosymmetric superconductors\cite{H.Tou_JPSJ_1995,H.Mukuda_PRL_2010}. 
The difference of the decrease of $K_{\rm spin}$ in the SC state is considered to be related with the strength of spin-orbit coupling interaction, and thus, a systematic Knight-shift study in HF superconductivity is required.

\section{Conclusion}
In conclusion, we have performed $^{63}$Cu-NMR/NQR measurements using $S$-type single-crystal CeCu$_2$Si$_2$ in order to investigate its SC and magnetic properties.
The temperature dependence of $1/T_1$ at zero field was almost identical to that in polycrystalline samples down to 130~mK but deviated downward below 120~mK.
The $1/T_1$ dependence in the SC state could be reproduced by the two-gap $s_{\pm}$-wave and the two-band $d$-wave.
Taking into account the recent results of the field-angle dependence of the specific heat, the two-gap $s_{\pm}$-wave model is plausible.
In magnetic fields, the spin susceptibility in both directions clearly decreased below $T_{\rm c}$.
The residual part of the spin susceptibility was well understood by the residual density of state arising from the vortex cores under a magnetic field.
Above $H_{c2}$, no obvious magnetic anomaly was observed in $S$-type CeCu$_2$Si$_2$ down to 150~mK, although the AFM fluctuations were enhanced on cooling. 
Thus, the present $S$-type single-crystal sample was in the paramagnetic state close to a quantum critical point, and superconductivity emarges out of the strong AFM fluctuations.  

\section*{Acknowledgments}
The authors acknowledge F. Steglich, S. Yonezawa, Y. Maeno, Y. Tokiwa, Y. Yanase, S. Shibauchi, H. Ikeda, Y. Matsuda, and Y. Kitaoka for fruitiful discussions. 
This work was partially supported by Kyoto Univ. LTM center, and Grant-in-Aids for Scientific Research (KAKENHI) (Grant Numbers JP15H05882, JP15H05884, JP15K21732, JP25220710, JP15H05745, and JP17K14339). 

\end{document}